\documentclass[pageno]{jpaper}
\pdfoutput=1

\usepackage[normalem]{ulem}
\usepackage{CJKutf8} % Remove
\usepackage{makecell}
\usepackage{algorithmicx}
\usepackage{algpseudocode}
\usepackage{algorithm}
\usepackage{pifont}

\begin{document}
\begin{CJK*}{UTF8}{bsmi} % Remove

\title{LIRS: Enabling efficient machine learning on NVM-based storage via a lightweight implementation of random shuffling}

\author{Zhi-Lin Ke \\ \small National Taiwan University \\ \small r05922083@ntu.edu.tw \and Hsiang-Yun Cheng \\ \small Academia Sinica \\ \small hycheng@citi.sinica.edu.tw \and Chia-Lin Yang \\ \small National Taiwan University \\ \small yangc@csie.ntu.edu.tw }

\date{}
\maketitle

\thispagestyle{empty}

\begin{abstract}

Machine learning algorithms, such as Support Vector Machine (SVM) and Deep Neural Network (DNN), have gained a lot of interests recently.
When training a machine learning algorithm, randomly shuffle all the training data can improve the testing accuracy and boost the convergence rate.
Nevertheless, realizing training data random shuffling in a real system is not a straightforward process due to the slow random accesses in hard disk drive (HDD).
To avoid frequent random disk access, the effect of random shuffling is often limited in existing approaches.
With the emerging non-volatile memory-based storage device, such as Intel Optane SSD, which provides fast random accesses, we propose a lightweight implementation of random shuffling (LIRS) to randomly shuffle the indexes of the entire training dataset, and the selected training instances are directly accessed from the storage and packed into batches.
Experimental results show that LIRS can reduce the total training time of SVM and DNN by 49.9\% and 43.5\% on average, and improve the final testing accuracy on DNN by 1.01\%.

\end{abstract}

\section{Introduction}

Machine learning algorithms have recently grown in prominence as they can provide effective solutions for many classification and regression tasks, including computer vision, speech recognition, and natural language processing.
A machine learning model needs to be trained by a massive number of training instances before it can be utilized for an inference task.
In practice, mini-batch gradient descent~\cite{li2014efficient} that updates the model based on an average of the gradients inside each subset (refer to a batch) of training instances is usually used for training machine learning models, including the popular Support Vector Machine (SVM) and Deep Neural Network (DNN). 

For the machine learning algorithms that use mini-batch gradient descent for training, the way to organize batches greatly impact the training efficiency.
Randomly shuffling all the training instances to form batches can improve the testing accuracy and boost the convergence rate~\cite{bengio2012practical,bottou2009curiously, ioffe2015batch,montavon1998tricks}.
Nevertheless, realizing training data random shuffling in a real system is not a straightforward process due to slow random accesses in hard disk drive (HDD).
To avoid frequent random disk access, existing approaches, such as the block minimization framework (BMF)~\cite{BMF} of SVM and the TensorFlow input pipeline (TFIP)~\cite{tensorflow2015-whitepaper, tfapi14} of DNN, often sacrifice some degree of randomness when shuffling the training data.
As random disk access is slow, these existing designs usually perform only one full-range training data shuffling before the training starts and only partially shuffle the training data during the training process.
In addition, the shuffling process often occupies large memory space.

In recent years, non-volatile memory based storage devices, such as Intel Optane SSD~\cite{optane_p4800x}, have attracted a lot of interests and have emerged as a promising storage solution for future computer systems.
These new storage devices provide fast random accesses.
For example, Intel Optane SSD can achieve 550,000 IOPs on random reads, about 916x and 1.28x higher than the throughput of HDD and NAND Flash SSD random reads.
With faster random accesses, we can re-think how to achieve random shuffling effectively by taking advantage of these new types of storage devices.

To exploit the fast random accesses offered by the new types of storage devices, we design a new shuffling method, a Lightweight Implementation of Random Shuffling (LIRS), to realize fully random shuffling.
The main idea of LIRS is to randomly shuffle the indexes of the entire training dataset and the selected training instances are directly accessed from the storage and packed into batches.
As I/O system calls rather than load/store instructions are used to implement random accesses, two techniques, Data Format Aware Location Generator and Page-aware Random Shuffling, are proposed to tackle the challenges of sparse data format and small training instance size. 

The new shuffling method, LIRS, has the following three advantages. 
First, this method can converge faster and achieve higher testing accuracy compared to the existing implementations.
Second, the initial data pre-processing stage is no longer needed.
Third, it does not occupy large memory resources as current random shuffling implementations.

In this paper, we evaluate the convergence rate, testing accuracy, and training time of existing solutions, SVM's BMF and DNN's TFIP, and compare them against our LIRS methodology.
The experiments are conducted on real machines with three different types of storage devices, including HDD, NAND Flash SSD, and Intel Optane SSD.
Results show that LIRS can significantly reduce total training time by increasing convergence rate and reducing initial data pre-processing time.
On average, LIRS can reduce total training time by 49.9\% for SVM and 43.5\% for DNN, compared to applying existing solutions on HDD.
The testing accuracy of DNN can be improved by 1.01\% as LIRS can increase the variation within and between batches to avoid stucking at local minimum.

\paragraph{We make the following contributions:}

\begin{itemize}
\item To our best knowledge, this is the first work to analyze the impact of different shuffling methods on convergence rate and training time, considering the access performance of different storage types.
\item We propose a novel data shuffling method, LIRS, which can achieve fully random data shuffling with minimal memory space requirements. LIRS can be applied on any machine learning algorithms that use mini-batch gradient descent for training, including the popular SVM and DNN.
\item We conduct comprehensive experiments to evaluate the total training time of different combinations of random shuffling methods and storage devices. Results show that simply replacing HDD by a faster storage device is not enough to get the optimal training time and designing a new shuffling method, such as LIRS, is necessary and beneficial. 
\end{itemize}

\section{Background}

In this section, we first overview the basics of machine learning training and the training process of two representative machine learning algorithms, Support Vector Machine (SVM) and Deep Neural Network (DNN). 
We then elaborate the importance of random shuffling in the training process, in terms of convergence rate and testing accuracy. 

\subsection{Basics of Machine Learning Training}

Machine learning enables computers to automatically make prediction or decision by training a machine learning algorithm (refer to model) to learn from the training dataset.
The training process includes three key steps~\cite{Goodfellow-et-al-2016}.
First, the model reads an input training instance (This refers to a single training material of total dataset, also called example or sample), which is composed of representative features, and derive a predicted value based on the statistical estimation.
Then, a loss function is utilized to compare the predicted value with the known target.
Finally, the model is updated based on the loss value, in order to derive a better model that can minimize the loss.
In practice, mini-batch gradient descent~\cite{li2014efficient} is commonly used to stably minimize the loss value by updating the model based on an average of the gradients inside each subset (refer to a batch) of input training instances.
These three steps are applied on all the input training instances in the training dataset, and training the entire training dataset once is called an {\bf epoch}.
This process repeats until the model reaches the convergence condition.
After training, we can use a testing dataset to evaluate how well a machine learning algorithm performs by measuring the testing accuracy.

There are two mainstream machine learning models, Support Vector Machine (SVM) and Deep Neural Network (DNN), which commonly use mini-batch learning and are very useful in large-scale classification and regression analysis problems. 

\noindent{\textbf{Support Vector Machine (SVM).}}
SVM is widely used for classification.  It separates classes with a linear function (called hyperplane). 
The training process of SVM is to find the largest-margin separating hyperplane, finally turn into minimize the objective function(also called cost function, loss function).
If the dataset can't fit into the memeory, the training process slows down due to frequent page faults.  
To solve this issue, Block Minimization Framework (BMF)~\cite{BMF} that uses LIBLINEAR~\cite{LIBLINEAR} as the optimizer is the state-of-the-art solution.
BMF splits the training dataset into batches and trains a batch at a time.
The batch size is usually set to the maximum possible size that can fit into main memory~\cite{BMF}.
SVM training is usually running on CPU instead of GPU, because LIBNEAR uses dual coordinate descent~\cite{coordinate_descent} to minimize objective value and the sequential design of coordinate descent makes it difficult to be parallelized.

\noindent{\textbf{Deep Neural Network (DNN).}}
Deep Neural network uses multiple layers of non-linear processing units for feature extraction and is a promising solution for many machine learning problems.
Common DNN architectures consist of convolutional layers, pooling layers, and fully-connected layers. After the operation of these layers, the model output the final prediction value. The training process is to find the best weights to correct prediction.
DNN models are typically trained using Stochastic Gradient Descent (SGD). 
Training data are randomly sampled into batches. 
Batches are fed into the model to traverse the model in two phases: forward and backward passes. 
The forward pass generates predictions, and calculating the loss between the prediction and the ground truth.
Then the backward pass backpropagates errors to obtain gradients to update model weights. 
DNN training is computing intensive, and is commonly accelerated with GPUs~\cite{kayidperformance, raina2009large, dean2012large}.
When training DNN model with GPU, the training data first be read from the storage device to the host main memory. 
After a series of data preprocessing, input training instances in host main memory are packed into batches and these batches are sent to GPU’s off-chip memory.
GPU then begins to train the DNN model after receiving these batches of training data.
Note that, when GPU training on a batch, the CPU side can continuously read data from the storage device. 
Thus the training can be more efficient because of the overlapping of computing and loading.

\subsection{Importance of Training Data Random Shuffling}

For mini-batch based machine learning, the way to organize batches greatly impact the effectiveness and efficiency of the algorithm.
Randomly shuffling training data to form batches can improve both testing accuracy and convergence rate~\cite{ ioffe2015batch,montavon1998tricks,bengio2012practical} by avoiding overfitting and preventing the training process from stucking at a local minimum~\cite{ioffe2015batch}.
If the order of input training instances within each epoch is the same, the model may use this input pattern as a way of reducing the training loss, which will be more likely to induce overfitting resulting in worse testing accuracy.
Figure~\ref{Fig: queue_size_impact} shows the impact of random shuffling when training with randomly shuffled data the testing accuracy can improve about 2\%$\sim$3\% compared to training with unshuffled data (The experiment setting is described in Section \ref{Neural Network - TensorFlow Input Pipeline}).
In addition, random shuffling can increase the convergence rate, ~\cite{bottou2009curiously} shows that training with random shuffle can make the loss dropping more faster.
In summary, randomly shuffle the training instances before each epoch is important when training with mini-batch learning, and which can achieve better testing accuracy and improve the convergence rate.

\section{Motivation}
\label{Challenge and Motivation}

Realizing training data random shuffling in a real system is not a straightforward process due to slow random accesses in HDD. 
To avoid frequent random disk access, the effect of random shuffling is often limited. With the emerging non-volatile memory-based storage device, such as Intel Optane SSD, which provides fast random accesses, we can re-think how to achieve random shuffling effectively by making good use of this new type of storage devices. In this section, we first introduce the current random shuffling implementations in SVM and DNN. We then explain the opportunity a  non-volatile memory-based storage device can offer. 

\subsection{SVM - Block Minimization Framework}
\label{SVM - Block Minimization Framework}

\begin{figure} 
\centering 
\includegraphics[width=0.4\textwidth]{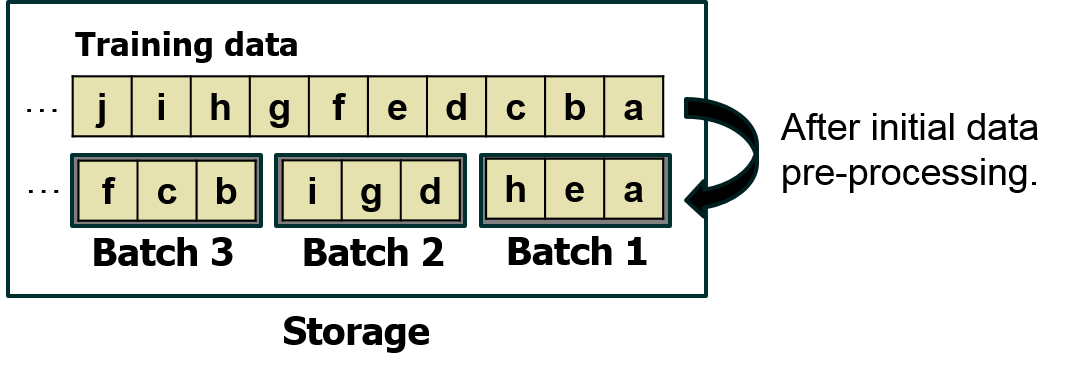} 
\caption{Initial splitting process of SVM.} 
\label{Fig: Initial Splitting Process}
\end{figure}

Block Minimization Framework (BMF)~\cite{BMF} is the state-of-the-art SVM training framework designed for large-scale training. BMF performs a full-range training data shuffling before starting the training process. 
The initial data pre-processing stage sequentially reads training instances to the main memory, and then write each training instance back to a randomly selected file (batch). The instances of the same batch are stored in contiguous chunks.
As demonstrated in Figure~\ref{Fig: Initial Splitting Process}, instance d, e, f are placed adjacently in the original training data files. After initial shuffling, they are assigned to different batches (batch 2, batch 1, and batch 3). 
In this way, when reading a batch from the storage, it achieves the random effect via sequential disk accesses.  
This shuffling process is quite time-consuming since it reads the entire training set and write training data back to the storage randomly. Therefore, it is only performed once. 
During the training process, BMF only reshuffles the accessing order of different batches at each epoch, while the training instances in each batch remain the same.

This shuffling implementation has two sources of inefficiency. First, even though the initial shuffling process is only performed once, it still takes a significant part of the training time(about more than 10\% total training time in mainstream SVM workloads~\cite{BMF}), as HDD random write is slow and SVM model is relatively simple. Second, at each epoch, it does not perform full-range random shuffling, i.e., the training instances of a batch do not change. This results in slower convergence. 
\vspace{-1ex}

\subsection{DNN - TensorFlow Input Pipeline}
\label{Neural Network - TensorFlow Input Pipeline}

\vspace{-2ex}

\begin{figure} 
\centering 
\includegraphics[width=0.48\textwidth]{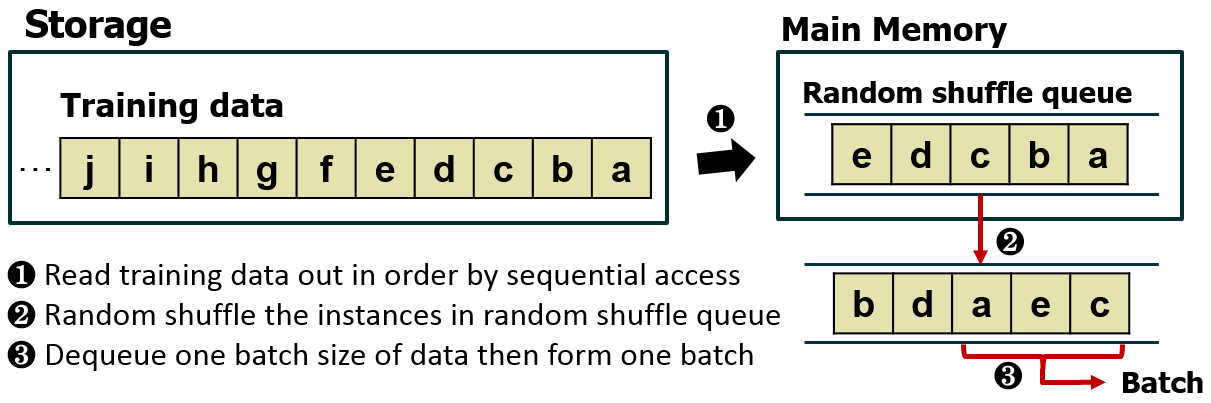} 
\caption{TensorFlow input pipeline.} 
\label{Fig: TensorFlow Input Pipeline}
\end{figure}

TensorFlow input pipeline~\cite{tensorflow2015-whitepaper, tfapi14} is an importing data scheme for retrieving training data. Similar to SVM, before training, an initial pre-processing is needed to completely shuffle the entire training dataset.
During training, to avoid random HDD accesses, TensorFlow input pipeline first sequentially loads portions of training dataset into the random shuffle queue in CPU memory, and then random shuffle the training instances stored in the random shuffle queue then dequeue a batch size of data to form a batch as demonstrated in Figure~\ref{Fig: TensorFlow Input Pipeline}. We can see that in this implementation, the shuffling degree is limited by the size of the memory queue. In the example shown in Figure~\ref{Fig: TensorFlow Input Pipeline}, the queue size is 5, so only instance a, b, c, d, e could be possibly assigned to the same batch. Therefore, for instances that are not placed closely to each other, they have nearly no chances to be packed into the same batch. Therefore, the random degree is limited. 

\begin{figure}
\subfloat[AlexNet] {  
\includegraphics[width=0.24\textwidth]{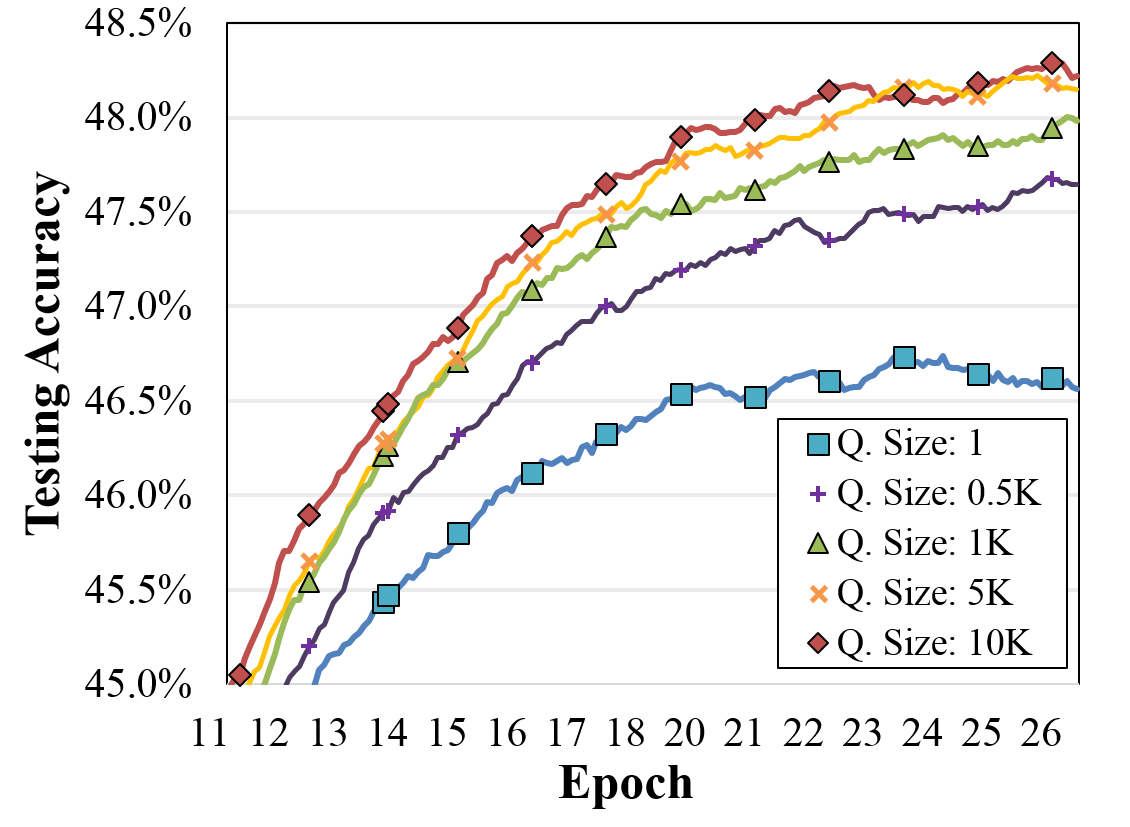}
\label{AlexNet}
}
\subfloat[OverFeat] { 
\includegraphics[width=0.24\textwidth]{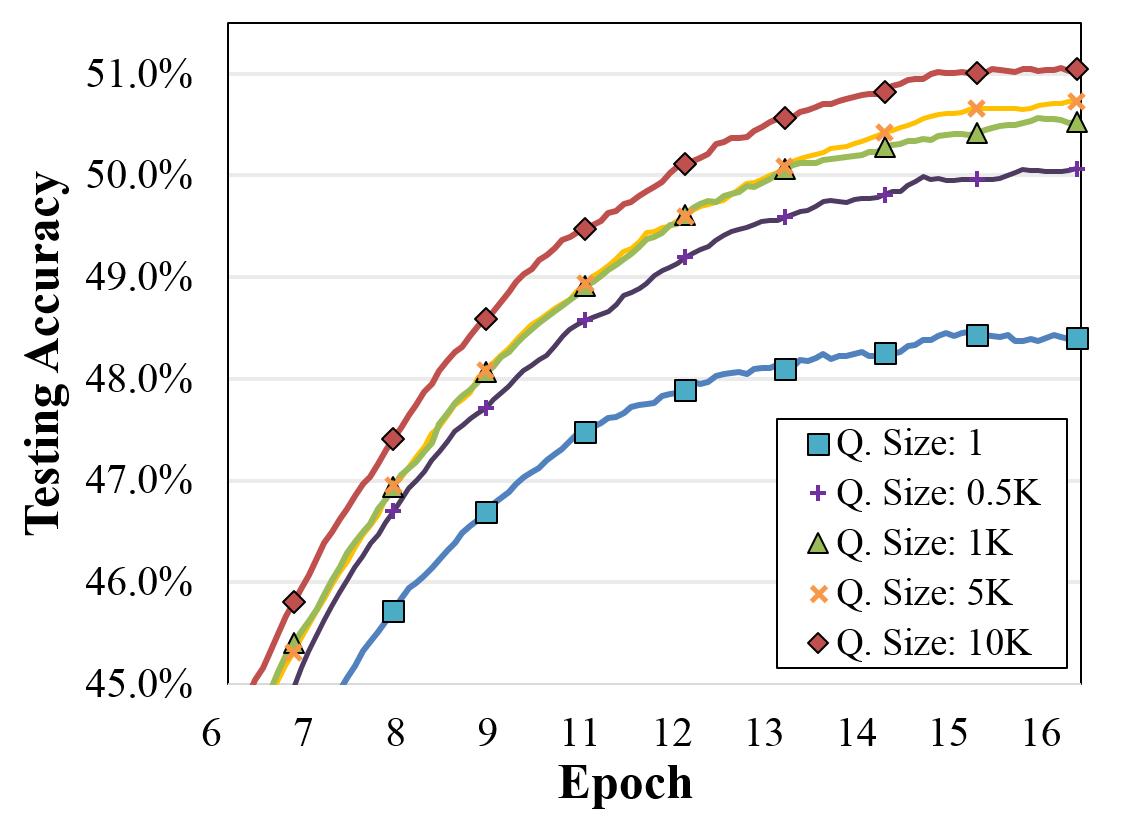}
\label{OverFeat}
}
\caption{The impact of queue size on testing accuracy.}
\label{Fig: queue_size_impact} 
\end{figure}

Figure~\ref{Fig: queue_size_impact} shows the impact of queue size on the testing accuracy for using  when using ImageNet~\cite{imagenet_cvpr09} to train AlexNet~\cite{alexnet} and OverFeat~\cite{sermanet2013overfeat}.
As illustrated in the figure, training with larger queue size can achieve better testing accuracy for both AlexNet and OverFeat.
When the queue size equals to 10000, AlexNet and OverFeat can achieve 48.3\% and 51.1\% testing accuracy, much higher than the case with queue size equals to 1 (i.e., no random shuffling). 
Although setting a large queue size can improve convergence rate and testing accuracy, a large random shuffle queue would occupy a significant amount of memory space.
For example, for ImageNet, each photo has 196608 features and each feature is represented by a 32-bit floating point number, the random shuffle queue with 10000 training instances occupies about 7.3GB memory.

In summary, the random shuffling implementation of TensorFlow input pipeline is inefficient, because (1) initial pre-processing is needed to completely shuffle the entire dataset and (2) the random degree of the shuffling is limited by the size of random shuffle queue in CPU main memory. Larger queue size can improve testing accuracy but cost tremendous amount of memory space.

\subsection{Opportunity for Efficient Random Shuffling with NVM-based Storage}

\vspace{-2ex}

In recent years, non-volatile memory based storage devices, such as Intel Optane SSD, have been envisioned as an efficient storage solution for future computer systems. These new storage devices offer fast random read and write.   
As shown in Table~\ref{storage}, the random read throughput of a modern HDD~\cite{WD10EZEX} is only 600 IOPS, about 67x lower than the throughput of sequential reads.
On the contrary, a modern NAND Flash SSD~\cite{intel750} can achieve 430,000 IOPS when executing random reads, much better than HDD random reads and even better than HDD sequential reads.
Moreover, Intel Optane SSD~\cite{optane_p4800x} which employed 3D Xpoint technology~\cite{3dxpoint} to improve performance can achieve 550,000 IOPS on random reads, only slightly lower than its sequential read throughput.

\begin{figure} 
\centering 
\includegraphics[width=0.3\textwidth]{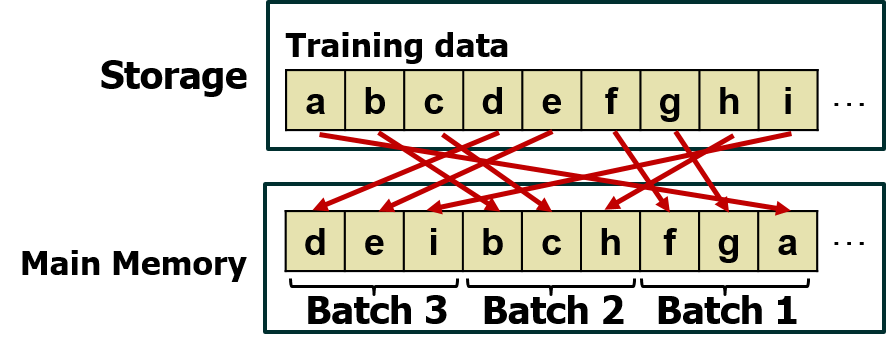} 
\caption{Shuffling data by random accessing on storage.}
\label{Fig: shuflle_by_random_access}
\end{figure}

With better random access performance in advanced SSDs, we can design new shuffling methods to realize fully random shuffling as demonstrated in Figure~\ref{Fig: shuflle_by_random_access}.  The indexes (instance ID) of the entire training data are randomly shuffled, selected training instances are accessed directly from the SSD and packed into batches.
The new shuffling method has the following three advantages.
{\bf First, at each training epoch, we can perform full-range random shuffling so compared to the existing implementations, this method can converge faster and achieve higher testing accuracy.  Second, the initial data pre-processing stage is no longer needed.  Third, it does not occupy large memory resources as current random shuffling implementations}.

Traditionally, when HDD with slow random access speed is employed during small-scale training (i.e., the entire training dataset can fit into the main memory), all the training instances would be sequentially loaded into the main memory then do shuffling.
Then, fully random shuffling is achieved by randomly shuffling the index (instance ID) of all the training instances and packing the training instances into batches.
If the underlying storage devices, such as Intel Optane SSD, can provide better random access performance, we can also shuffle training instances by directly random accessing the specific data.

\section{Random Shuffling with SSD}

\subsection{Lightweight Implementation of Random Shuffling (LIRS)}
\label{Shuffliing Design with SSD - LIRS}

\begin{figure} 
\centering 
\includegraphics[width=0.48\textwidth]{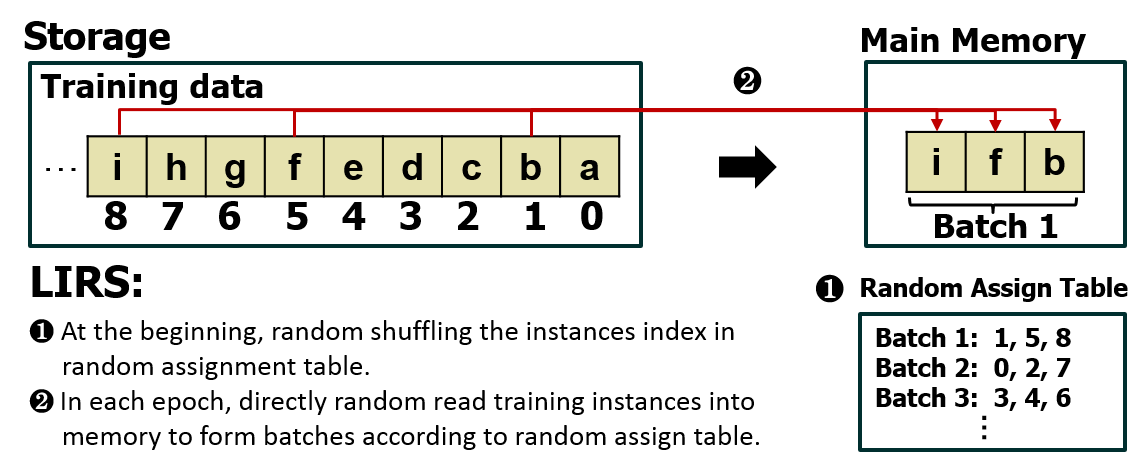} 
\caption{LIRS}
\label{Fig: LIRS}
\end{figure}

To exploit the fast random accesses of advanced storage devices, such as Intel Optane SSD, we propose LIRS, a Lightweight Implementation of fully Random Shuffling. 
LIRS achieves fully random shuffling by directly reading the training instances from the storage in random order to form batches at every epoch.
In the following, we explain the main idea of LIRS and how to solve the design challenges introduced by directly random access training instances from the storage.

\paragraph{LIRS design}
The core concept of LIRS is to randomly assign the training instances to each different batches on the host side to achieve the random shuffling effect.
LIRS maintain a random assignment table in the main memory, which recording the instance assigned information of each batch.
We will give each training instance a unique ID, and the shuffling process just random assign the IDs to different batches and record on the random assignment table.
Note that, LIRS will re-assign the training instances before each epoch to achieve completely random shuffling effect.
When the training starts, LIRS will query the random assignment table to get the instances ID which assign to the batches, and directly create random access to read out those instances.
As shown in Figure~\ref{Fig: LIRS}, instance ID 1, 5, 8 was assigned to batch 1 in this epoch.
When batch 1 needs to be read, LIRS will create random access to read out these instances directly.
When all batches have been read for a round, the random assignment table will be reshuffled again.
Thus, we can achieve fully random shuffling at every epoch using only small memory space (random assignment table plus one batch size).

\paragraph{Challenges}

To realize fully random shuffling by directly and randomly accessing training instances from storage, two challenges need to be solved as I/O system calls rather than regular load/store instructions are used to implement random accesses.
First, in order to randomly read any specific instance by an I/O system call, we must know the position of each training instance in the file.
The position of each instance can be calculated by the file start position plus the offset of the instance.
If the lengths of each instance are the same, we can simply multiply the size of each instance by the index of the instance to get the offset.
However, if the data is stored in sparse format, which only saves non-zero feature values, we cannot directly obtain the offset by simple arithmetic as the lengths of each instance are different.
Second, when the size of an instance is smaller than the size of an OS virtual page, using I/O system calls to get each instance may cause inefficient memory space utilization.
Since a standard I/O system call read/write one page of data at a time, an entire page of data would be loaded into main memory from the storage even when only part of the loaded page (i.e., the randomly selected training instance) is accessed. 
Most of the other instances reside in the same page as the selected instance would not be used before the page is being evicted, because the entire training dataset is too large to fit into the main memory and the randomly selected instances have poor spatial locality. 
The inefficient page utilization wastes precious memory space and induces redundant page transfers between the main memory and storage.

\begin{figure} 
\centering 
\includegraphics[width=0.4\textwidth]{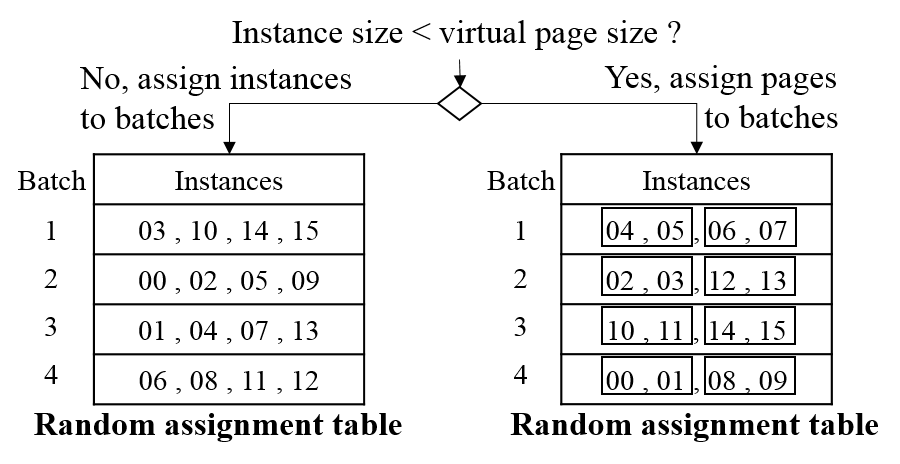} 
\caption{Example of using instance and page as the minimum assigning unit.} 
\label{Fig: Example of assigning data}
\end{figure}

\paragraph{Data Format Aware Location Generator}

To tackle the first challenge, we propose to use a data format aware location generator to obtain the location of each training instance according to the stored data format.
When the training instances are stored in non-sparse format, i.e., the lengths of each instance are the same, the offset of a training instance can be directly obtained by multiplying the instance ID with instance size.
On the other hand, if the training instances are stored in sparse format, i.e., only non-zero feature values are stored and the lengths of each instance are different, the location generator scans the entire training dataset and record the offset of each instance in an offset table during the initial pre-processing stage.
Then, during the training process, the offset of a training instance is obtained by accessing the offset table.
Based on the offset derived by the location generator, we can randomly read any specific instance by an I/O system call regardless of the data format.

\paragraph{Page-aware Random Shuffling}

To address the second challenge, we propose a page-aware random shuffling scheme that can help to improve memory utilization and reduce redundant page transfer when the size of a training instance is smaller than the size of an OS virtual page.
Different from the naïve instance-based random shuffling that only uses part of the loaded page (i.e., the randomly selected training instance) when training a batch, our page-aware random shuffling uses a page as the minimum random shuffling unit and group the training instances within the same page into the same batch. 
This approach can fully use the loaded page and prevent redundant page transfer that happens when a page is evicted before all the training instances within the page is trained.

Figure~\ref{Fig: Example of assigning data} shows an example that randomly partition 16 training instances into 4 batches based on the page-aware random shuffling scheme.
If the size of a training instance is larger than a page, we use instance as the random shuffling unit to randomly group four instances into a batch when generating the random assignment table, as shown in Figure~\ref{Fig: Example of assigning data}(a).
On the contrary, if the size of a training instance is smaller than a page, the training instances within the same page, such as instance 4 and 5, are randomly assigned to the same batch, as shown in Figure~\ref{Fig: Example of assigning data}(b).
In this example, the size of a training instance is only half of a page size, so using page-aware random shuffling can save up to 50\% page transfer between the storage and main memory, resulting in shorter data loading time.
Note that page-aware random shuffling may sacrifice some random degree and slightly decrease the convergence rate, as training instances within the same page are forced to be grouped into the same batch.
Nevertheless, the benefit of significant loading time reduction usually outweighs the minimal increase in convergence time, as will be shown in our experiments in Section~\ref{Page assignment vs. Instance assignment}.

\subsection{Memory usage analysis}
The required memory resources for LIRS is to maintain the random assign table and the offset table if training data are kept in the sparse format. The random assign table only records instance IDs, which could be represented in 4 to 8 bytes considering training data sizes. Take ImageNet as an example, there are a total of 1281167 instances, and ID could be represented with 4 bytes. Therefore, the total size of the table is only about 4.89MB. The memory capacity of today's server is usually in the range of 32GB. The memory overhead of the random assign table is quite low. For the offset table,  we only stores the offset of each instance, which could also be represented in 4 to 8 bytes. Similarly, the memory overhead of the offset table is also low.

\subsection{Comparison with conventional approaches}
\label{Comparison with conventional approach}
In this section, we compare the training process of LIRS with conventional methods designed for HDD and discuss the advantages provided by LIRS.

\begin{figure} 
\centering 
\includegraphics[width=0.48\textwidth]{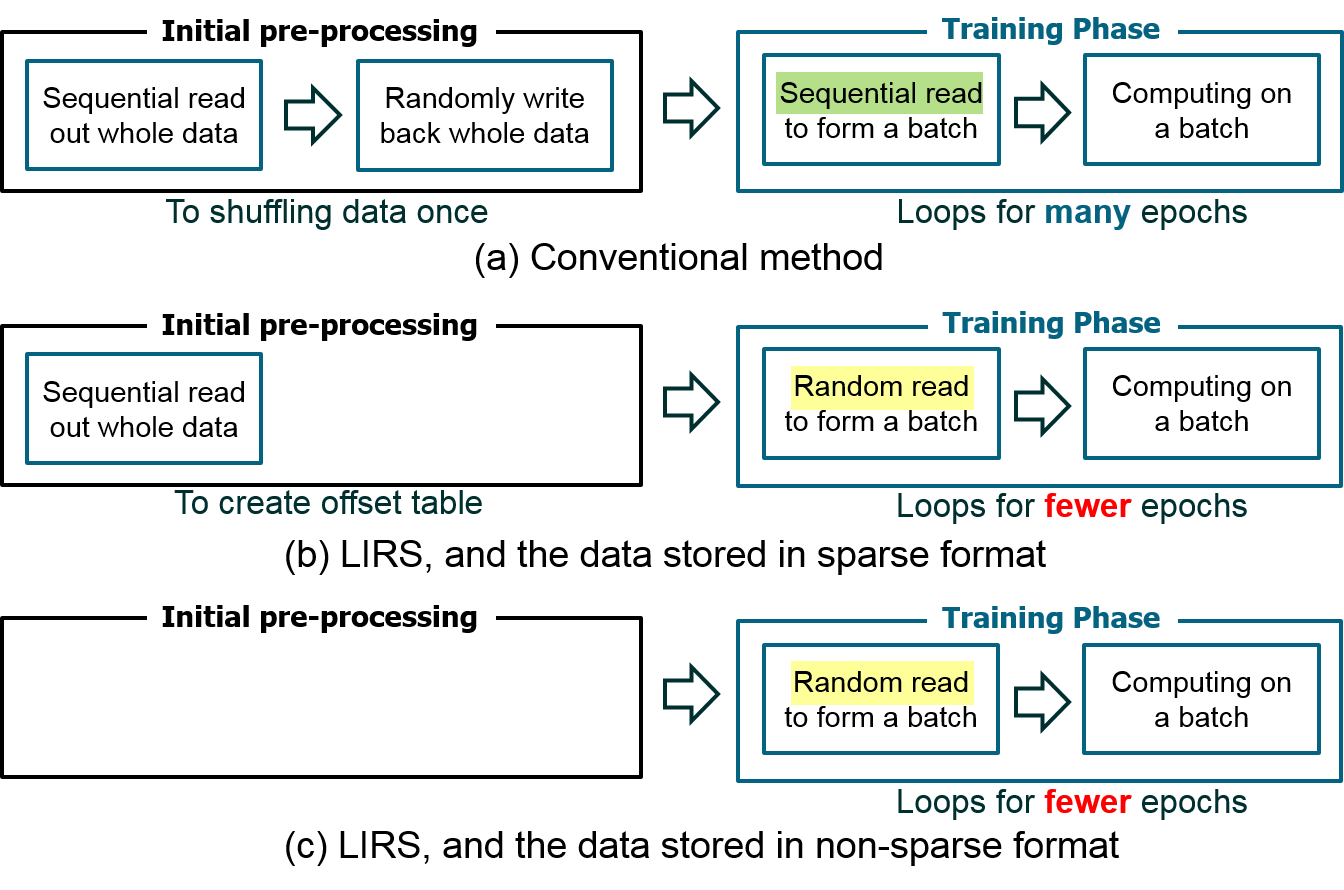} 
\caption{LIRS vs. Conventional method - SVM} 
\label{comparison_SVM}
\end{figure}

\paragraph{SVM}

Figure~\ref{comparison_SVM} shows the training flow of SVM when different shuffling methods are applied. 
For the initial pre-processing stage, when the conventional BMF (Figure~\ref{comparison_SVM}a) is applied, all the training instances need to be shuffled once before the training starts.
The shuffling requires sequentially reading out all the training instances from storage and then randomly writing back the shuffled instances to storage. 
Using LIRS can reduce the initial pre-processing time and the amount of savings depends on the data format.
When the training instances are stored in the sparse format (Figure~\ref{comparison_SVM}b), there is no need to shuffle data during the pre-processing stage but the entire training dataset need to be sequentially read out once to record the location of each training instance in the offset table.
Comparing to conventional BMF, LIRS can save the random write back time at the pre-processing stage with sparse data format.
On the other hand, when the training instances are stored in non-sparse format (Figure~\ref{comparison_SVM}c), the initial pre-processing can be completely eliminated compared to the conventional approach.

For the training phase, conventional BMF gets a batch from storage by performing sequential reads (Figure \ref{comparison_SVM}a), 
Since BMF only partially shuffles training instances, the training phase loops for many epochs to reach the convergence criteria.
In contrast, LIRS randomly access the storage to form batches (Figure \ref{comparison_SVM}b,\ref{comparison_SVM}c).
As the training instances are shuffled thoroughly, it takes less epochs to converge than conventional BMF.

The downside of LIRS is that it needs to perform randomly reads, which is a bit slower than sequential reads used in conventional BMF. For SVM, it is hard to overlap fetching training data with computation due to large batch sizes.   
Thus, when LIRS is applied, the impact on training time greatly depends on the random read performance of the underlying storage. If the random reads of the underlying storage are fast enough, such as when using Intel Optane SSD as storage, the significant performance gain derived by fewer training epochs can outweigh the negative impact brought by the increased loading time.

\begin{figure} 
\centering 
\includegraphics[width=0.48\textwidth]{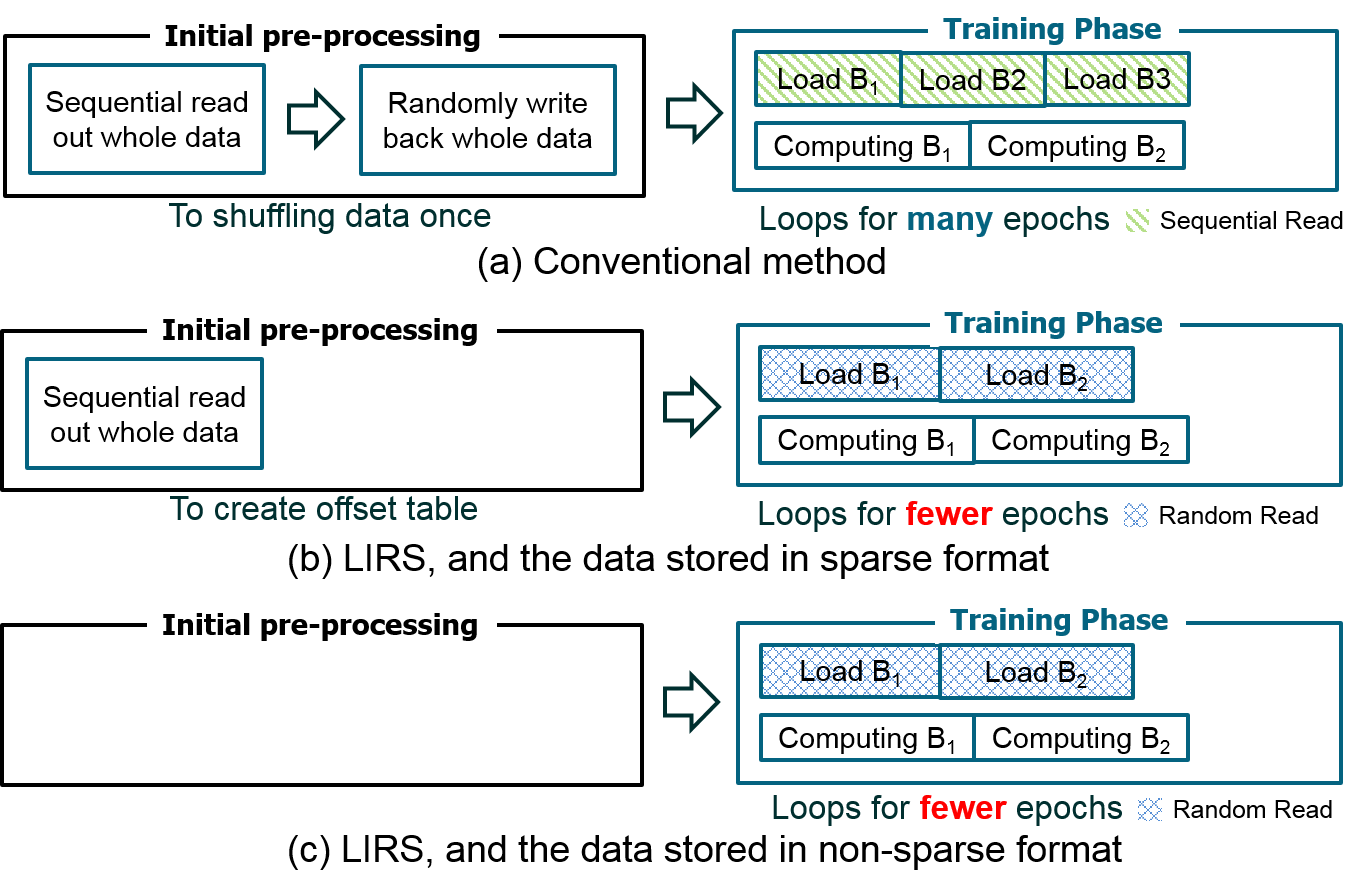} 
\caption{LIRS vs. Conventional method - DNN} 
\label{comparison_NN}
\end{figure}

\paragraph{DNN}

Figure~\ref{comparison_NN} shows the flow of DNN training when different shuffling methods are applied.
Simiar to SVM, LIRS can reduce the fime of the initial pre-processing stage. 
Since DNN has a much larger model than SVM, the initial pre-processing time is not as dominating as in SVM. 
However, due to ever-increasing dataset sizes, the pre-processing time is still quite long, from  several minutes to hours. Eliminating the data pre-processing stage frees CPUs for other usages, which is important in data centers.

For the training phase, as discussed before, LIRS could reduce the required epochs since it converges faster than the conventional TensorFlow input pipeline implementations. Similar to SVM, LIRS fetches training data via random read while the conventinal methods use sequential reads. However, unlike SVM, in DDN, the loading time and computation time can be overlapped, as CPU can prefetch the next batch from storage at the same time when GPU is computing a batch.
The computation time of a batch is usually longer than the loading time, as DNN models are generally complex.
Thus, the performance of LIRS is not sensitive to the random access speed of SSD, as long as the random access speed is not too slow.

\section{Evaluation}

In this section, we evaluate the convergence rate, testing accuracy, and training time of conventional shuffling methods applied on SVM and DNN, and compare them against our LIRS methodology.
Results on systems with three different types of storage devices, including HDD, NAND Flash SSD, and Intel Optane SSD, are analyzed and discussed. 

\subsection{Experimental Setup}

\begin{table}
\scalebox{0.63}{
\begin{tabular*}{13.3cm}{lllcccl}
\hline
\multirow{2}{*}Dataset & Number of & Number of & total size & Avg Instance size & is\_sparse & Model\\
& instances & features & (GB) & (Bytes) &\\
\hline
webspam&200000&16609143&8.3&44560&Y&SVM \\ 
epsilon&400000&2000&8.9&24000&N&SVM  \\
kdd&19264097&29890095&6.5&362&Y&SVM  \\
higgs&10500000&28&3.2&327&N&SVM  \\
ImageNet&1281167&196608&234.6&196608&N&DNN  \\
\hline
\end{tabular*}}
\caption{Training datasets. Avg instance size $=$ total size / number of instances.}
\label{dataset}
\end{table}

We use different evaluation platforms for SVM and DNN experiments.
For SVM, we use a single-CPU system to conduct our convergence and performance studies.
The single-CPU system has a AMD A10-7850K processor with 32MB DDR3 memory.
Since we target on training large-scale SVM that cannot store the entire training dataset in main memory, we follow the method utilized by~\cite{BMF} and use Linux built-in cgroups~\cite{cgroups} to limit the maximum memory usage to 1GB.
For DNN, we conduct the experiments on NVIDIA's GTX 1070~\cite{geforce1070}, equipped with 8GB DDR5 memory.
Both SVM and DNN evaluation platforms are connected to three types of storage devices: conventional HDD (HDD-WD10EZEX~\cite{WD10EZEX}), SSD (SSD-Intel-750~\cite{intel750}), and Optane SSD (OptaneSSD-P4800x~\cite{optane_p4800x}), as listed in Table~\ref{storage}.
The random access throughput of SSD is much better than conventional HDD, and Optane SSD can provide the highest random access throughput among these three storage devices.

\paragraph{Workloads}

For SVM, we use four classification datasets, as shown in Table~\ref{dataset}, to evaluate the impact of different shuffling methods on convergence and performance.
Among these four datasets, webspam~\cite{webspam} and kdd~\cite{yu2010feature} are stored in sparse binary format, while epsilon~\cite{yuan2012improved} and higgs~\cite{lichman1990uci} are stored in non-sparse binary format.
The average size of training instances is smaller than OS virtual page size (4KB) at kdd and higgs, while the instance size of the other two datasets are larger. 
When implementing SVM, we follow the block minimization framework ~\cite{BMF} and use LIBLINEAR \cite{LIBLINEAR} as the optimizer. 
The batch size is set to the maximum possible size that can fit into main memory, as larger batch size enables faster convergence~\cite{BMF,serafini2005working}.
Thus, we partition the training dataset of webspam, epsilon, kdd, and higgs into 40, 30, 40 and 40 batches respectively.

For DNN, we use ImageNet~\cite{imagenet_cvpr09}, which contains 1000 classes of images, to train three different DNN models, including AlexNet~\cite{alexnet}, OverFeat~\cite{sermanet2013overfeat} and VGG16~\cite{vgg16}.
These models are well known and all have achieved outstanding results in the ILSVRC~\cite{ILSVRC15} competition.
We use Tensorflow r1.4 \cite{tfapi14} framework with a deep learning library, cuDNN 5.1 \cite{chetlur2014cudnn}, to implement these three DNN models.
The batch size is set to 128, 128, and 32 for AlexNet, OverFeat, and VGG16.

\begin{table}
\centering
\scalebox{0.6}{
\begin{tabular*}{14cm}{lcc}
\hline
Storage&sequential read/write(IOPS)&random read/write(IOPS) \\
\hline
WD 10EZEX (HDD)&40000/36000&600/300\\
INTEL® SSD 750&563000/230000&430000/230000\\
INTEL® OPTANE™ SSD DC P4800X&614000/512000&550000/500000\\
\hline
\end{tabular*}}
\caption{Evaluated storage devices.}
\label{storage}
\end{table}

\paragraph{Random shuffling methods}

In this paper, we implement different shuffling methods to study their impact on model convergence and training performance.
For SVM, we implement and evaluate block minimization framework (BMF) introduced in Section~\ref{SVM - Block Minimization Framework} and our LIRS.
For DNN, we implement and evaluate TensorFlow input pipeline (TFIP) explained in Section~\ref{Neural Network - TensorFlow Input Pipeline} and our LIRS.
The default random shuffle queue size in TFIP is set to 10000 instances. 

\paragraph{Evaluation metrics}

To analyze the impact of different random shuffling methods, we use the following three metrics for evaluation: convergence rate, testing accuracy, and  total training time.
For SVM, we use the descending rate of the relative function value difference~\cite{BMF} to represent the convergence rate, and for DNN, we use the descending rate of the validation loss to represent the convergence rate.
The total training time includes the time spent on pre-processing stage and training phase, and can be calculated by the following equation:

\vspace{-2ex}

\begin{small}
\begin{equation}\label{total training time}
\ T_{Total}= T_{preprocess} + (T_{load} + T_{comp} - T_{overlapping}) * \#Epochs
\end{equation}
\end{small}

\vspace{-2ex}

, where $T_{preprocess}$ is the pre-processing time, $T_{load}$ is the time spent on loading data from storage and decoding the data into memory objects, $T_{comp}$ is the time spent on computing, $T_{overlapping}$ is the overlapped time between loading and computing, and $\#Epochs$ represents the number of epochs spent on training the model until reaching the targeted convergence level\footnote{The targeted convergence level is set to the minimum training loss when baseline BMF or TFIP is applied.}

\subsection{Experimental Results of SVM}
\subsubsection{Convergence Rate and Testing Accuracy}

\begin{figure}
\subfloat[webspam (Obj)] {  
\includegraphics[width=0.24\textwidth]{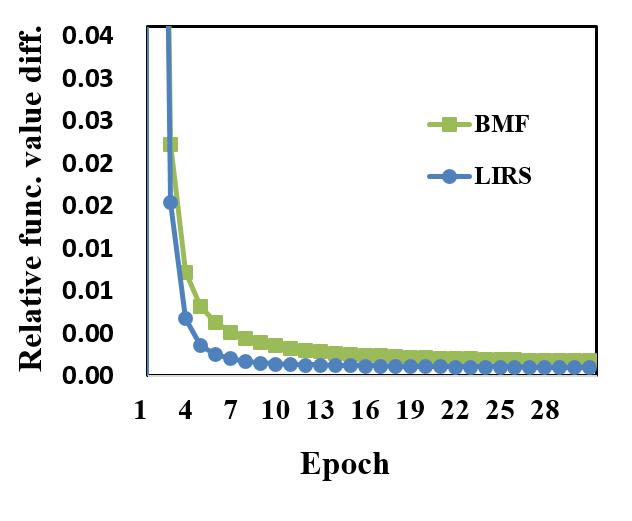}
\label{web_obj}
}
\subfloat[epsilon (Obj.)] { 
\includegraphics[width=0.24\textwidth]{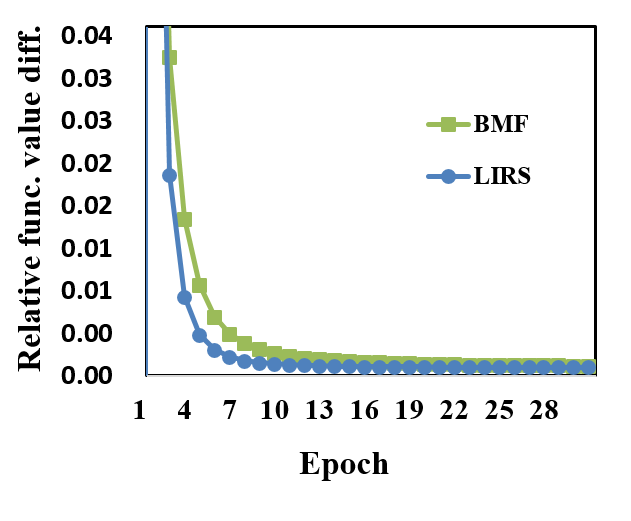}
\label{epsilon_obj}
}
\newline
\newline
\subfloat[kdd (Obj.)] { 
\includegraphics[width=0.24\textwidth]{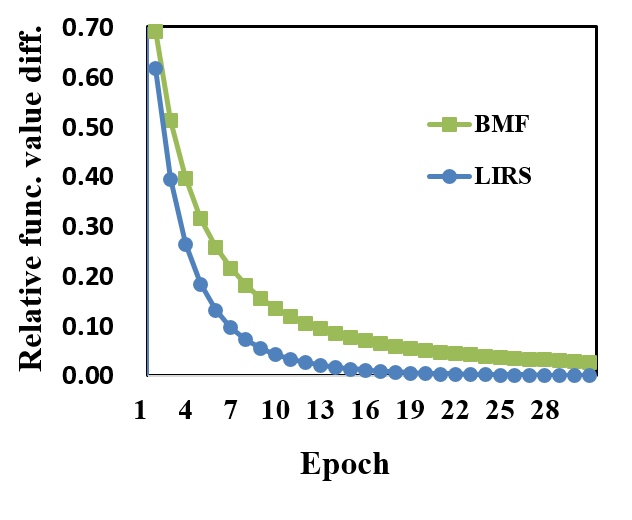}
\label{kdd_obj}
}
\subfloat[higgs (Obj.)] { 
\includegraphics[width=0.24\textwidth]{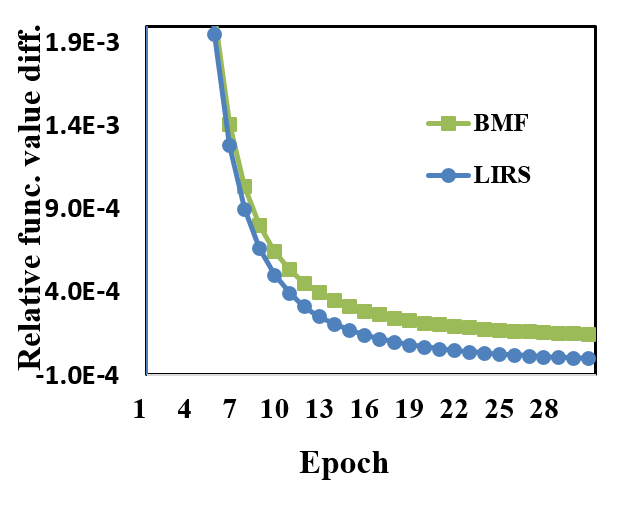}
\label{higgs_obj}
}
\caption{Relative function value difference of SVM training when different shuffling methods are applied.} 
\label{svm_loss_accuracy} 
\end{figure}

Figure~\ref{svm_loss_accuracy} shows the relative function value difference at different epochs when BMF and LIRS are applied. 
Greater slope at relative function value difference indicates faster convergence rate.
As shown in Figure~\ref{svm_loss_accuracy}, LIRS converges faster than BMF at all of the four training datasets, since BMF only partially shuffles the training instances and keeps the same set of training instances within each batch at every epoch.
With faster convergence rate, LIRS can spend fewer epochs to achieve the same relative function value difference as BMF, as shown in Table~\ref{comparison epoch}.
While BMF needs to train for 30 epochs to converge to its minimum relative function value difference, LIRS only needs to train webspam, epsilon, kdd, and higgs for 7, 12, 11, and 17 epochs respectively to achieve the same convergence level.
Although LIRS can provide higher degree of randomness than BMF during shuffling, the testing accuracy of BMF and LIRS are similar and the difference is within 0.15\%, as shown in Table~\ref{SVM Best Testing Accuracy}.
Since the objective function of LIBLINEAR optimizer is linear and there is only one global optimal solution~\cite{hsieh2014divide, abdiansah2015time}, it is less likely to stuck at local minima even if the training instances are not thoroughly shuffled.

\begin{table}
\centering
\scalebox{0.8}{
\begin{tabular*}{8cm}{lcc|lcc}
\hline
Dataset&BMF&LIRS&Dataset&BMF&LIRS\\
\hline
webspam&30&7&kdd&30&11\\
epsilon&30&12&higgs&30&17\\
\hline
\end{tabular*}}
\caption{Number of required SVM training epochs to achieve BMF's minimum relative function value difference.}
\label{comparison epoch}
\end{table}

\begin{table}
\centering
\scalebox{0.8}{
\begin{tabular*}{8cm}{lccc}
\hline
Dataset&BMF&LIRS&Improvement\\
\hline
webspam&99.20\%&99.19\%&-0.01\%\\
epsilon&89.80\%&89.73\%&-0.01\%\\
kdd&88.70\%&88.84\%&+0.14\%\\
higgs&64.29\%&64.23\%&-0.06\%\\
\hline
\end{tabular*}}
\caption{Testing accuracy of different datasets.}
\label{SVM Best Testing Accuracy}
\end{table}

\subsubsection{Total Training Time}

\begin{figure*} 
\includegraphics[width=1\textwidth]{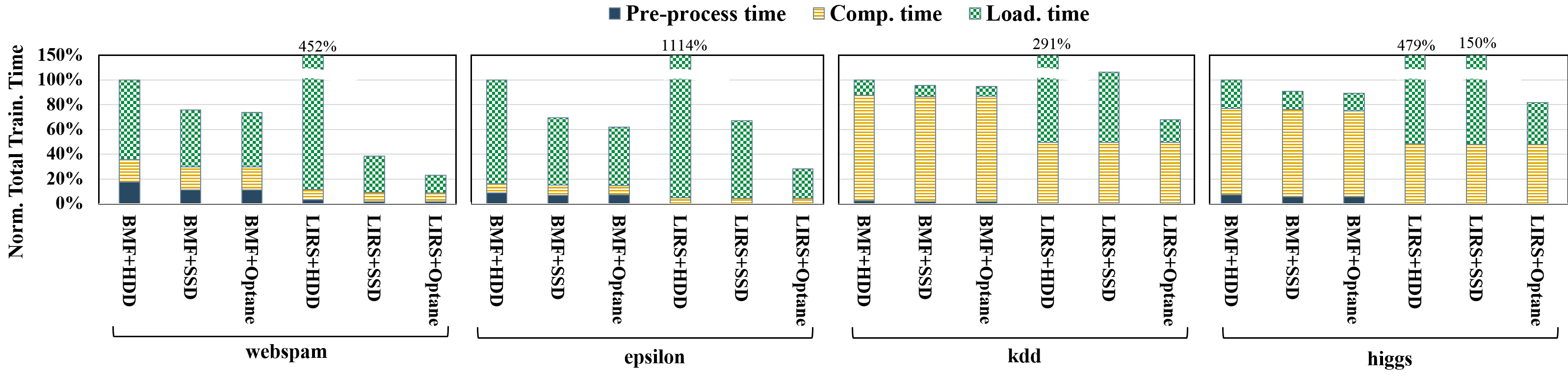} 
\caption{Total training time.} 
\label{Fig: performance_SVM}
\end{figure*}

Figure~\ref{Fig: performance_SVM} shows the total training time of different combinations of random shuffling methods and storage devices, normalized to BMF+HDD.
For fair comparison, we train the SVM model to the same level of relative function value difference, as listed in Table~\ref{comparison epoch}, to calculate total training time.

We first analyze the total training time of BMF when using different storage devices.
As shown in Figure~\ref{Fig: performance_SVM}, BMF benefits from the reduced loading time on SSD at all of the four datasets, because the sequential read throughput of SSD is much higher than HDD.
Replacing SSD by Optane SSD only slightly reduces the total training time of BMF, as most of BMF's storage accesses are sequential reads and Optane SSD provides only slightly better sequential read throughput than SSD.
Although simply replacing HDD by Optane SSD can help BMF to reduce the initial pre-processing time and loading time, the computing time and the number of training epochs remain the same.
Thus, to further reduce the total training time, a better training method, such as LIRS, is needed to reduce the computing time by increasing the convergence rate.

Figure~\ref{Fig: performance_SVM} shows that the random access performance of the underlying storage greatly impacts the total training time of LIRS.
LIRS randomly loads training instances from storage to achieve fully random shuffling, so the loading time greatly increases when LIRS is applied, especially when the underlying storage is HDD.
Thus, on average, the total training time of LIRS+HDD is 5.8x longer than BMF+HDD.
As the advance in storage technologies enables faster random reads and writes, the performance benefits provided by LIRS becomes noticeable.
Figure~\ref{Fig: performance_SVM} shows that, LIRS+SSD spends shorter total training time than BMF+SSD at webspam and epsilon, since (a) LIRS can reduce the initial pre-processing time at sparse webspam and completely eliminate the initial pre-processing time at non-sparse epsilon, and (b) LIRS fully shuffle the training instances at each epoch to achieve faster convergence.
However, when training the datasets with smaller instance size (kdd and higgs), the total training time of LIRS+SSD is longer than BMF+SSD, because LIRS generates significant amount of random reads and the random read speed of the evaluated SSD is not fast enough.
From the figure, we can also observe that LIRS+Optane can reduce the total training time by 76.9\%, 71.9\%, 32.2\%, and 18.4\% at webspam, epsilon, kdd, and higgs respectively, compared to BMF+HDD.
Note that, with Optane SSD's high random read throughput, LIRS can significant reduce the total training time at kdd and higgs even though severe amount of random accesses are generated.
In addition, LIRS+Optane can provide significant performance improvement compared to BMF+Optane, indicating that designing LIRS to cope with the advance in storage technologies is necessary and beneficial.

\subsubsection{Page-aware Random Shuffling vs. Instance-based Random Shuffling}
\label{Page assignment vs. Instance assignment}

\begin{figure}
\subfloat[Total training time] {  
\includegraphics[width=0.24\textwidth]{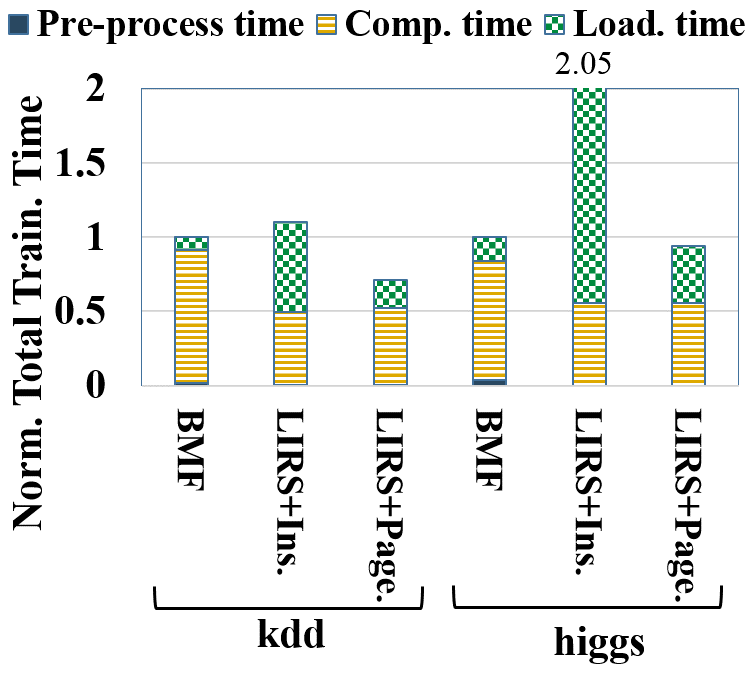}
\label{Fig: page_vs_ins}
}
\subfloat[Avg. comp. time and load. time per epoch] { 
\includegraphics[width=0.24\textwidth]{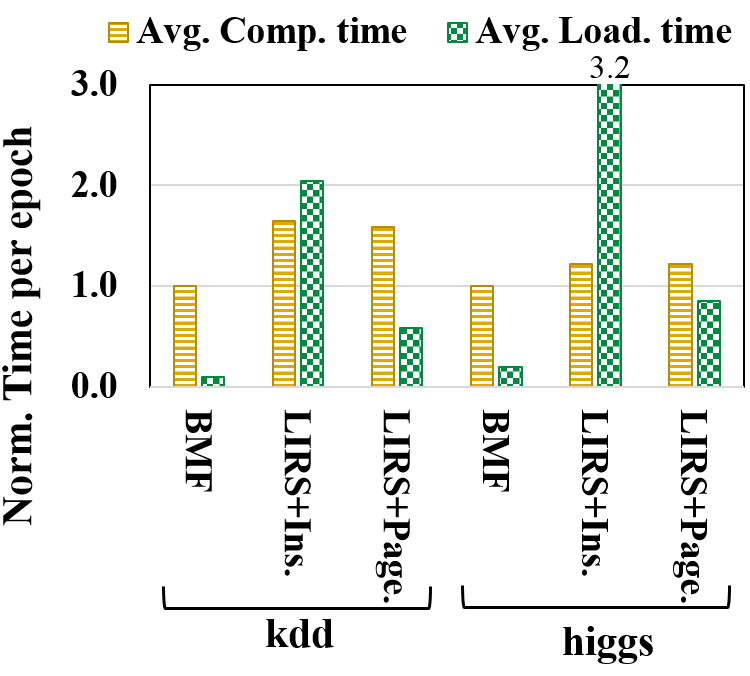}
\label{Fig: page_vs_ins_avg}
} 
\caption{Comparison between instance-based random shuffling and page-aware random shuffling.}
\label{nn_loss_accuracy} 
\end{figure}

Figure~\ref{Fig: page_vs_ins} shows the total training time of instance-based random shuffling (LIRS+Ins.) and page-aware random shuffling (LIRS+Page) normalized to BMF when the size of each training instance is smaller than an OS virtual page. 
As shown in the figure, LIRS+Inc. greatly increases the loading time as many instances within the same page are not utilized before being evicted , and the reloading of these instances causes redundant page loading.
On the contrary, LIRS+Page uses a page as the minimum random shuffling unit and groups the training instances within the same page into the same batch, in order to better utilize the loaded page and prevent redundant page transfer between the storage and main memory.
Thus, as shown in Figure~\ref{Fig: page_vs_ins}, LIRS+Page reduces the loading time by 68.7\% and 73.5\% at kdd and higgs, compared to LIRS+Inc..
As a result, LIRS+Inc. performs worse than BMF on Optane SSD while LIRS+Page can reduce the total training time by 28.6\% and 6\% over BMF, even though LIRS+Page sacrifices some degree of randomness during shuffling and slightly increases the number of training epochs (1 and 0 additional epochs at kdd and higgs).

Although LIRS+Page can greatly reduce the loading time compared to LIRS+Ins., the loading time of LIRS+Page is still much longer (about 2x) than BMF.
The reason is that the training instances are not page-aligned and part of the last instance in a page may reside on the next page.
When selecting a page of instances into a batch, two read system calls are generated to load two pages (one selected page plus one additional page) from the storage, in order to get the last instance that cross the boundary of a page.
Since LIRS+Page randomly selects pages of instance, other instances reside on the additional page may not be accessed before the page is evicted.
Thus, at the worst case, each page will be loaded twice, resulting in 2x increase in loading time.
We will solve this issue in our future work.

\subsubsection{Overhead Analysis}

Table~\ref{LIRS memory overhead} lists LIRS’s hardware overhead.
LIRS needs an additional random assignment table to record which instance belongs to which batch.
The size of the random assignment table is equal to the number of instances multiplied by the size of instance ID (8 bytes).
In addition to random assignment table, if the training instances are stored in sparse format, an offset table is required.
The size of the offset table is equal to the number of instances multiplied by the size of offset (8 bytes).
As shown in the table, LIRS introduces less than 1\% memory space overhead for webspam and epsilon in a 1GB main memory.
The memory overhead of kdd and higgs are higher because these two datasets contain a significant number of instances.

\begin{table}
\centering
\scalebox{0.75}{
\begin{tabular*}{10cm}{lcccc}
\hline
&webspam&epsilon&kdd&higgs\\
\hline
Random Assign Table&1.53MB&3.05MB&147MB&80.11MB\\
Offset Table&1.53MB&0&147MB&0\\
\hline
\end{tabular*}}
\caption{LIRS memory overhead}
\label{LIRS memory overhead}
\end{table}

\subsection{Experimental Results of DNN}
\subsubsection{Convergence Rate and Testing Accuracy}
\label{The impact of shuffling to convergence}

\begin{figure}
\subfloat[AlexNet] {  
\includegraphics[width=0.15\textwidth]{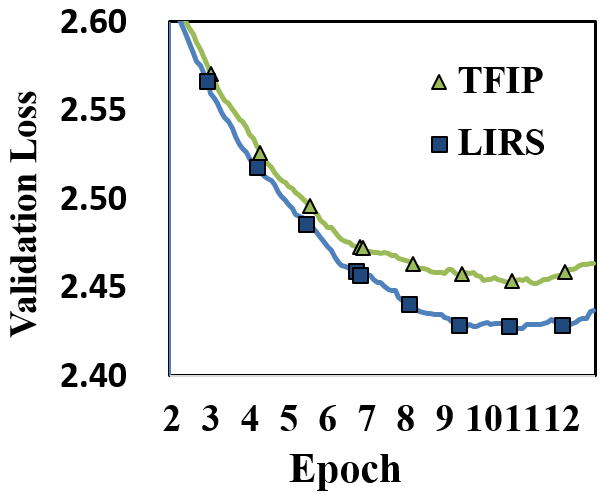}
\label{loss_alexnet}
}
\subfloat[OverFeat] { 
\includegraphics[width=0.15\textwidth]{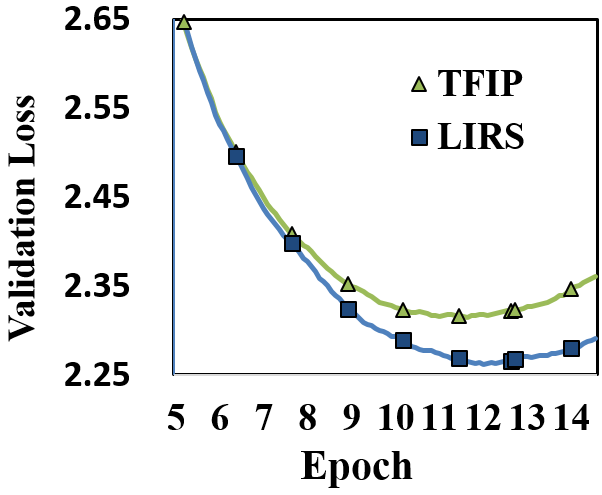}
\label{loss_overfeat}
}
\subfloat[VGG16] { 
\includegraphics[width=0.15\textwidth]{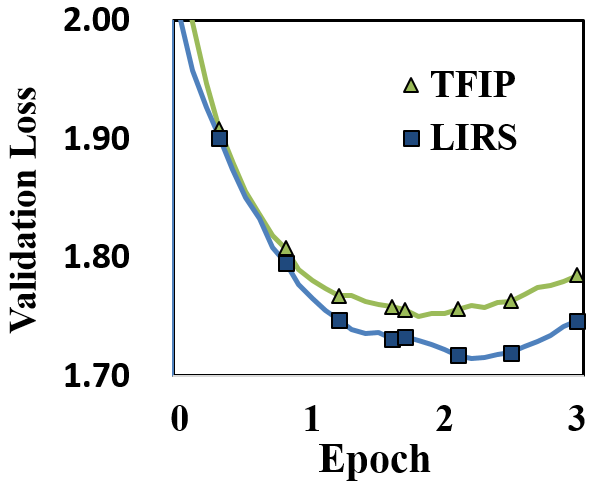} 
\label{loss_vgg16}
} 
\caption{Validation loss of DNN training when different shuffling methods are applied.} 
\label{nn_loss} 
\end{figure}

Figure~\ref{nn_loss} shows the validation loss at different epochs when TFIP and LIRS are applied.
Greater slope at validation loss indicates faster convergence rate.
As shown in Figure~\ref{nn_loss}, LIRS converges faster than TFIP when training all the three DNN models, since the degree of random shuffling is limited by the size of the random shuffle queue when TFIP is applied.
With faster convergence rate, LIRS can spend fewer epochs to achieve the same validation loss as TFIP, as shown in Table~\ref{comparison epoch DNN}
For example, TFIP needs to train AlexNet for 17.5 epochs to converge to its minimum validation loss, while LIRS only needs to train for 13.6 epochs to achieve the same convergence level.
In addition to the improvement of convergence rate, LIRS also helps to improve testing accuracy, as shown Figure~\ref{NN Best Testing Accuracy}.
Since LIRS can increase the variation within and between batches by thoroughly shuffling the training instances, the training process is less likely to stuck at local minima.
Thus, the testing accuracy of AlexNet, OverFeat, and VGG16 are increased by 0.65\%, 0.86\%, and 1.51\% respectively when LIRS is applied, compared to conventional TFIP.

\begin{table}
\centering
\scalebox{0.8}{
\begin{tabular*}{5.2cm}{lcc}
\hline
Model&TFIP&LIRS\\
\hline
AlexNet&17.5&13.6\\
OverFeat&11.9&9.4\\
VGG16&2.1&1.6\\
\hline
\end{tabular*}}
\caption{Number of required DNN training epochs to achieve TFIP's minimum validation loss.}
\label{comparison epoch DNN}
\end{table}

\begin{table}
\centering
\scalebox{0.8}{
\begin{tabular*}{7.2cm}{lccc}
\hline
Model&TFIP&LIRS&Improvement\\
\hline
AlexNet&48.29\%&48.94\%&+0.65\%\\
OverFeat&51.09\%&51.95\%&+0.86\%\\
VGG16&62.39\%&63.90\%&+1.51\%\\
\hline
\end{tabular*}}
\caption{Testing accuracy of different models on DNN}
\label{NN Best Testing Accuracy}
\end{table}

\subsubsection{Total Training Time}

\begin{figure*} 
\includegraphics[width=1\textwidth]{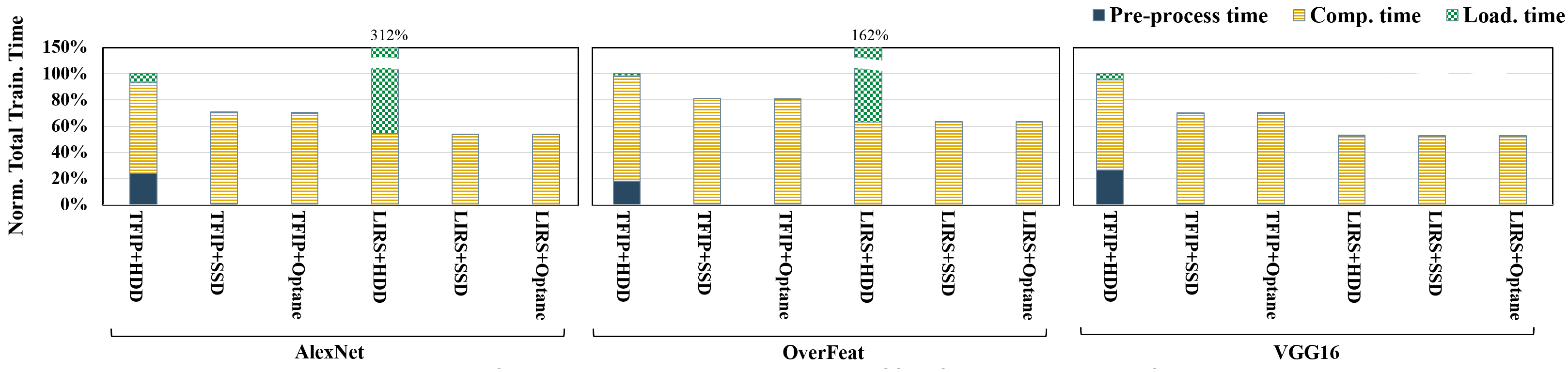} 
\caption{Total training time of DNN.} 
\label{Fig: performance_NN}
\end{figure*}

Figure~\ref{Fig: performance_NN} shows the total training time of different combinations of random shuffling methods and storage devices, normalized to TFIP+HDD.
For fair comparison, we train the DNN models to the same level of validation loss, as listed in Table~\ref{comparison epoch DNN}, to get the total training time.

We first analyze the performance of TFIP when using different storage devices.
As shown in Figure~\ref{Fig: performance_NN}, using storage devices with shorter access latency can reduce the total training time when TFIP is applied.
The faster sequential read and random write speed provided by SSD can help TFIP to reduce the initial pre-processing time and loading time.
Thus, TFIP+SSD can reduce total training time by 29.04\%, 18.92\%, and 29.68\% at AlexNet, Overfeat, and VGG16 respectively, compared to TFIP+HDD.
The total training time of TFIP+Optane are similar to TFIP+SSD in all of the three DNN models, because the loading time and computing time can overlap each other and the loading time was totally hide by computing time.
Although simply replacing HDD by faster storage devices can help TFIP to reduce the initial pre-processing time and loading time, the computing time remains the same.
Since the total training time of DNN is dominated by the computing time, even if we have a storage device faster than Optane SSD, we cannot reduce the total training time further when TFIP is applied.
Thus, to further reduce the total training time, it is necessary to improve the training methods, such as using LIRS to increase the degree of randomness during shuffling.

Figure~\ref{Fig: performance_NN} shows that using LIRS to train DNN models can significantly reduce the total training time when SSD and Optane SSD are employed as the storage device.
Although using LIRS increases the loading time, the loading time can almost totally be hidden (overlapped) by the long computing time spent on training the complex DNN models.
Thus, LIRS can reduce the total training time on SSD and OptaneSSD by increasing the degree of randomness during shuffling to improve the convergence rate.
If the computing time is significantly longer than the loading time, such as when training the most complex VGG16 model, LIRS can even get performance improvements when the underlying storage device is HDD.
From the figure, we can also observe that LIRS+Optane performs similar to LIRS+SSD.
Since the loading time and computing time can be overlapped and the computing time is longer, the performance of LIRS is not sensitive to the random access speed of SSD, as long as the random access speed is not too slow.
In summary, LIRS+Optane can reduce total training time of AlexNet, OverFeat, and VGG16 by 46.15\%, 36.90\%, and 47.56\% respectively, compared to TFIP+HDD.
The performance improvement of LIRS+Optane is higher than TFIP+Optane, indicating that it is necessary to design LIRS to cope with the advance in storage technologies.

\subsubsection{Overhead Analysis}

LIRS needs 9.8MB (< 0.1\%) additional memory space to store the random assignment table when using ImageNet to train DNN models.
Since ImageNet is stored in non-sparse format, no offset table is required.
Comparing to TFIP that needs 7.3GB memory space to implement the random shuffle queue, LIRS can save a large amount of CPU memory space. 

\section{Related works}

\cite{meng2017convergence} compared the impact of different shuffling degrees on convergence, but they only discussed the case that training dataset can fit into memory. In real situation, the dataset is usually larger than the memory space, so the storage access must be considered. \cite{lim2016analysis} discussed the impact of training dataset stored in different file systems and database, focusing on the overhead caused by the OS kernel. They also did not study the impact of data shuffling and storage device access performance on overall training time.
 To our best knowledge, we are the first work to analyze the impact of shuffling on convergence and training time, considering the random access performance of different storage devices.

Exsisting approaches usually deal with memory limitations by using mini-batch learning. In SVM, BMF~\cite{BMF} is used to split the training dataset into batches and then load and compute batches sequentially. BMF is designed to reduce the HDD I/O overhead, but it is not the best design if the underlying storage offers faster random accesses. Later, SBM~\cite{chang2011selective} further reduces the amount of disk access by keeping informative instances in memory to speed up the convergence. However, the shuffling methods still follows the partially shuffled batch order. Using LIRS can further improve the random shuffle degree and increase the convergence rate.

\section{Conclusion}

For large-scale learning, existing shuffling methods sacrifice the degree of random shuffling to reduce random storage accesses. 
We propose LIRS, a Lightweight Implementation of Random Shuffling, which exploits the fast random read performance of NVM-based storage and can achieve fully random shuffling with small memory overhead. Due to the increase in the random shuffling degree, experimental results show that LIRS can significantly improve the convergence rate in both SVM and DNN. With higher convergence rate and shorter initial pre-processing time, the total training time is greatly reduced by 49.9\% in SVM and by 43.5\% in DNN on average. In addition, LIRS improves the testing accuracy of DNN by 1.01\%. Results also show that simply replacing HDD by a faster storage device is not enough to get the optimal training time and designing a new shuffling method, such as LIRS, is necessary and beneficial.

\bibliographystyle{ieeetr}
\bibliography{references}

\end{CJK*} % Remove
\end{document}